\documentclass[ aip,amsmath, reprint,amssymb,]{revtex4-1}

\usepackage{color}
\usepackage{xcolor}
\definecolor{mydarkblue}{rgb}{0.1,0,0.55}
\definecolor{darkgreen}{rgb}{0.1,0.5,0.1}
\definecolor{orange}{rgb}{1,0.4,0.0}
\usepackage{hyperref}
\usepackage{graphicx}
\usepackage{dcolumn}
\usepackage[utf8]{inputenc}
\usepackage[T1]{fontenc}
\usepackage{mathptmx}
\usepackage{amsfonts}
\usepackage{amsmath}

\def\ups{Universit{\'e} Paris-Saclay, CEA, Laboratoire {M}ati{\`e}re sous conditions extr{\^e}mes, 91680 Bruy{\`e}res-le-Ch{\^a}tel, France}
\def\cea{CEA-DAM-DIF, F-91297 Arpajon, France}
\def\corr{Corresponding author}

\newcommand{\udens}{\ensuremath{\,\text{g/cm}^3}}
\newcommand{\Abinit}{\ensuremath{\,\textsc{Abinit}}}
\newcommand{\Vasp}{\ensuremath{\,\textsc{Vasp}}}
\newcommand{\QE}{\ensuremath{\,\textsc{Quantum Espresso}}}
\newcommand{\ext}{{\tt extended FPMD} method}
\newcommand{\extshort}{{\tt extended FPMD}}
\newcommand{\extveryshort}{{\tt Ext}}
\newcommand{\G}{\ensuremath{{G}  }}
\newcommand{\e}{\ensuremath{{\bf e}  }}
\newcommand{\LLO}{LLO}
\newcommand{\etal}{{\it et al.}}

\begin{document}

\title{Requirements for very high temperature Kohn-Sham DFT simulations and how to bypass them}
\author{A. Blanchet}
\email [email address: ]{augustin.blanchet@cea.fr}
\affiliation{\cea}
\affiliation{\ups}

\author{M. Torrent}
\affiliation{\cea}
\affiliation{\ups}
\email [email address: ]{marc.torrent@cea.fr}

\author{J. Cl\'erouin }
\affiliation{\cea}
\affiliation{\ups}
\affiliation{\corr}
\email [email address: ]{jean.clerouin@cea.fr}

\date{\today}
\begin{abstract}
In  high temperature density functional theory simulations (from tens of eV to  keV)  the total number of  Kohn-Sham  orbitals  is a critical quantity to get accurate results. To establish the relationship between the number of orbitals and the level of occupation of the highest energy orbital, we derived a model based on the homogeneous electron gas properties at finite temperature. This model predicts the total number of orbitals required to reach a given level of occupation and thus a stipulated precision. Levels of occupation as low as $10^{-4}$, and below, must be considered  to get converged results better than 1\%, making high temperature simulations very time consuming beyond a few tens of eV. After  assessing the predictions of the model  against previous results and \Abinit\, minimizations, we show how the \ext\, of Zhang \etal\, [PoP {\bf 23} 042707, 2016]  allows to  bypass these strong constraints on the number of orbitals at high temperature.
\end{abstract}
\maketitle

\section{Introduction}
Kohn-Sham density functional theory (KSDFT \cite{HOHE64,KOHN65}) simulations are now a well-established technique to compute static and dynamical properties of matter at ambiant conditions, and are implemented in simulation packages such as \Vasp\, \cite{KRES96}, \QE\,\cite{GIAN09} or \Abinit  \cite{ABINIT,GONZ09,GONZ20}. 
The extension of this technique towards hot systems was historically introduced  in  \Vasp, through the Mermin finite temperature functional  \cite{MERM65}, where orbitals are populated according to the Fermi-Dirac statistics. This approach, well adapted up to a few eV, which  corresponds to the domain  of liquid metals \cite{KRES93},  becomes more expensive  at higher temperatures (tens of eV), entering the warm dense matter regime (WDM). More precisely, the shape of the Fermi-Dirac distribution, strongly depends on the ratio $\theta$ of the temperature to the Fermi temperature. The latter scales as $n_e^{2/3}$, where $n_e$ is the electronic density (see Eq.\,\ref{eq:fermi}). For materials at standard density, $\theta$  quickly becomes higher than one as the temperature rises, flattening  the Fermi distribution. This flattening is responsible for the large number of orbitals to be included in the simulations. Conversely, for strongly compressed materials, $\theta$ remains lower than one (degenerate matter), even at high temperatures. The Fermi-Dirac distribution stays close to its zero temperature shape, allowing for KSDFT simulations of compressed materials at high temperatures \cite{RECO09,SJOS16,DING17,HU20,BETH20,DING17,SOUB19,HU20} with a small number of orbitals.

To avoid this difficulty, techniques taking advantage of the principle  of the {\em nearsightedness}  of the one particle density matrix, introduced by Kohn \cite{KOHN96,PROD05},  have been developed. Among them, the path-integral Monte-Carlo (PIMC) method in the restricted path approximation, is now able to treat elements of the second row of the periodic table\cite{POLL84,ZHAN17,MILI15}.  Other methods based on the density matrix are also explored (spectral quadrature \cite{SURY18}, stochastic \cite{CYTT18}, mixed \cite{WHIT20}),  and are increasingly faster  with increasing temperature, scaling linearly with the size of the system. But these approaches are computationally very demanding, especially at low temperature, with the exception of the mixed method which takes advantage of both representations (orbitals and density matrix). Another alternative, much less time consuming,  is provided by average atom models, that solve the Schr\"odinger or the Dirac equations in a box, coupled or not with a structural evaluation through the resolution of integral equations such as hyper-netted-chain (HNC). One can mention Purgatorio \cite{STER07}, the pseudo-atom molecular dynamics model \cite{STAR16,STAR17}, the neutral pseudo-atom model \cite{PERR95} or the SCAALP model \cite{BLAN04, FAUS10}.

For the dense plasma regime, a simplified approach  has been proposed,  using an orbital-free (OFDFT) formulation of the electronic kinetic energy based on the finite temperature Thomas-Fermi  theory and its extensions \cite{LAMB07, LAMB13}. This method, well adapted for dense plasmas simulations and their mixtures \cite{TICK16, WHIT17,CLER20},  is very fast for temperatures beyond tens of eV, but yields poor results below. To ensure accurate results at low temperature with the OFDFT approach, and, in particular, a satisfying description of chemical bonds, more  elaborated  functionals and a better treatment of exchange-correlation must be included \cite{KARA12,KARA12b,KARA13,SOJS13,SOJS15}.  The transition from  the KSDFT formulation  at low temperature  to the OFDFT, as the temperature rises, is described in  \cite{MAZE07,SJOS15b}, and  solutions have been proposed using the bootstrap method \cite{SHEP14} to compute Hugoniot or by introducing a KSDFT  reference points \cite{DANE12} to build a coherent equation of state. 

More recently, Zhang \etal\, \cite{ZHAN16} proposed to extend the KSDFT approach at high temperature by replacing high energy orbitals by plane waves. Introducing the homogeneous electron gas (HEG) properties at high energy, the so-called \ext\, makes the connection with OFDFT approaches at high temperature but preserves the KS orbitals at low temperature.  By keeping a minimum number of orbitals as the temperature rises, this method allows for a continuous transition from cold materials to hot plasmas.

A systematic comparison of the different approaches mentioned above was the subject of two blind comparison workshops, one  on the equations of state \cite{GAFF18} and the other on transport coefficients \cite{GRAB20}.

 In this paper we provide quantitative prescriptions  for high temperature KSDFT simulations. After discussing an example illustrating the convergence of the pressure with the number of orbitals involved in the calculation, we introduce a simple model based on the homogeneous electron gas  to predict the number of orbitals required to reach a stipulated occupancy. We then compare these predictions with \Abinit\, minimizations and with previous  molecular dynamics simulations at high temperature \cite{RECO09,KARA12b,LUO20}.  To bypass these convergence constraints, we discuss the recently introduced \ext. We  show, through its implementation in the \Abinit\, software package, how  very high temperature (2\,000\,eV) simulations are easily performed with the same level of accuracy as PIMC calculations despite a low number of orbitals.

\begin{figure}[!t]
\begin{center}
\includegraphics[width=7cm]{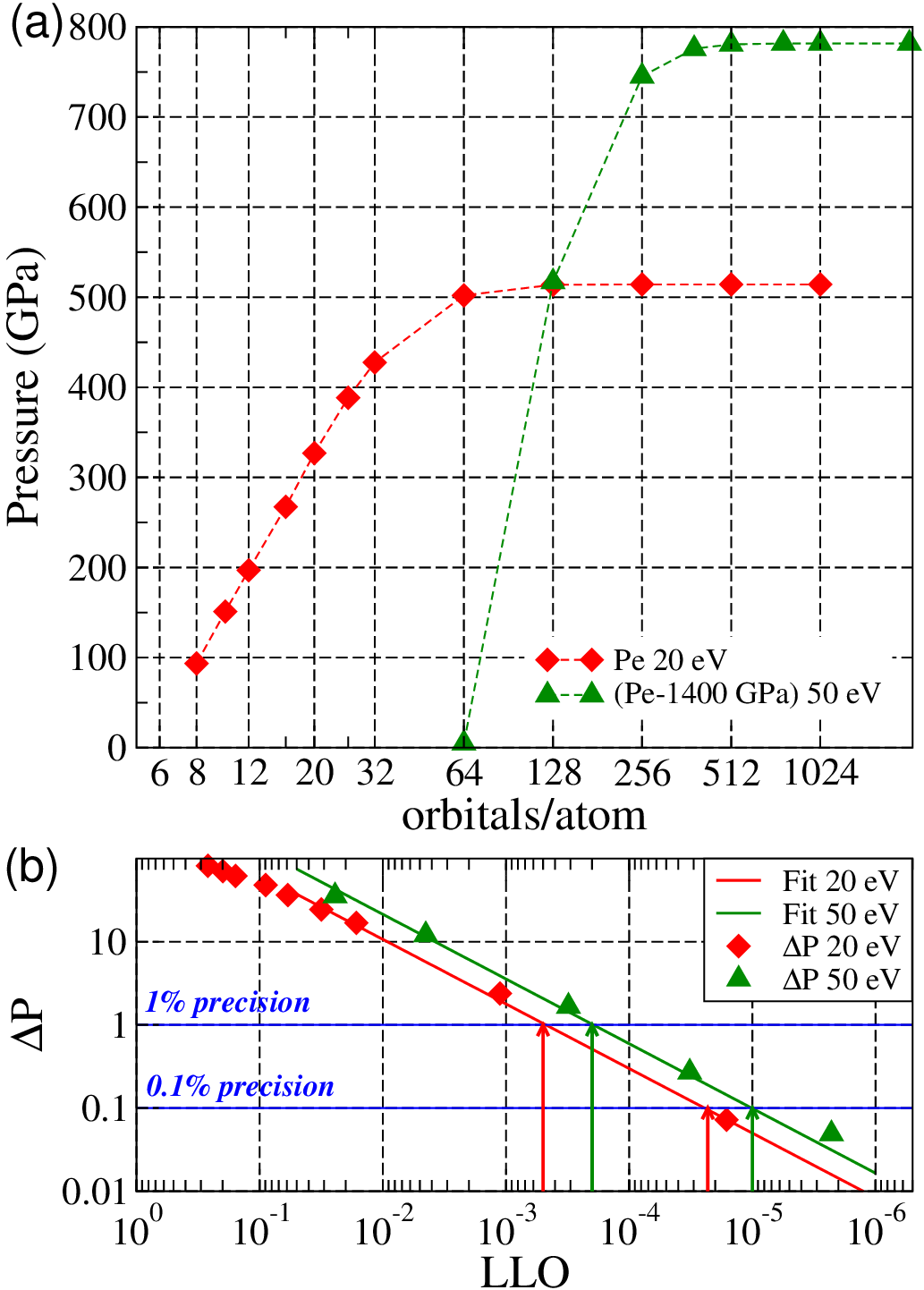}
\caption{(Color online) (a) Convergence of the electronic pressure versus the number of orbitals per atom  for a 4 aluminum atoms  at standard density  $\rho_0$ and at 20 (red diamonds) and  50\,eV (green triangles). For 50\,eV pressures  are shifted down by 1400\,GPa.
(b) Relative accuracy, in percentage, of the pressure calculation $\Delta P=100|P-P_0|/P_0$ versus LLO, where $P_0$ is the pressure given by the highest number of orbitals (see Tables\,\ref{table:occup} and \ref{table:occup2}). Red diamonds are for 20\,eV data, and green up triangles for 50\,eV. Corresponding straight lines are least squares fits and the horizontal blue lines the 1\% and 0.1~\% precision. Red and green arrows indicate the LLO required for 1\% (resp. 0.1\,\%) precision at 20 and 50\,eV. Note that the abscissa scale is reversed. Data with a LLO smaller than $10^{-6}$ are not represented.
 }
\label{fig:pabinit}
\end{center}
\end{figure}
\begin{table}[!t]
\caption{Finite temperature evaluation of  the electronic pressure of four aluminum atom at standard density and  20~eV  for a varying number of orbitals. The PAW pseudo-potential involves 11 electrons. The horizontal line between 64 and 128 orbitals delimits the 1\% precision region.}
\begin{center}
\begin{tabular}{|c|c|c|c|}
\hline \hline
Orbitals	&	\LLO		& 	Pressure	& $\Delta$P\\
/atom	&			&	GPa		&	\%	\\
\hline
\hline
8	&2.6 10$^{-1}$		&93.49901	&82 		\\
10	& 2.0 10$^{-1}$		&151.10035	&71 		\\
12	&1.6 10$^{-1}$		&196.87071	&62  		\\
16	&8.9	 10$^{-2}$		&267.26940	&48 		\\
20	&5.9 10$^{-2}$		&326.88243	&36 		\\
26	&3.2 10$^{-2}$		&388.38812	&24 		\\
32	&1.6 10$^{-2}$		&427.61530	&17 		\\
64	&1.1 10$^{-3}$		&501.95302	&2.4 		\\
\hline
128	&1.6 10$^{-5}$		&513.83618	&0.07		\\
256	&2.0 10$^{-8}$		&514.20696	&0.0002	\\
512	&3.5 10$^{-13}$	&514.20793	&	0.			\\
1024	&1.0 10$^{-15}$	&514.20793	&	0.			\\
\hline
\end{tabular}
\end{center}
\label{table:occup}
\caption{Same as Table~\ref{table:occup} at 50\,eV. The horizontal line between 256 and 384 orbitals delimits the 1\% precision region.}
\begin{center}
\begin{tabular}{|c|c|c|c|}
\hline \hline
Orbitals	&	\LLO		& 	Pressure	& $\Delta$P	\\
/atom	&			&	GPa		&	\%	\\
\hline
\hline
64	&2.4 10$^{-2}$	&1404.858	&35 	\\
128	&4.5 10$^{-3}$	&1917.023	&12 		\\
256	&3.1 10$^{-4}$	&2145.269	&1.7 		\\
\hline
384	&3.2 10$^{-5}$	&2175.835	&0.27			\\
512	&2.3 10$^{-6}$	&2180.651	&0.04	\\
768	&1.3 10$^{-7}$	&2181.678	&0.0017		\\
1024	&5.4 10$^{-9}$	&2181.722	&0.00027	\\
2048	& < 10$^{-18}$	&2181.716	&	0.			\\
\hline
\end{tabular}
\end{center}
\label{table:occup2}

\end{table}%

\section{Convergence with the number of orbitals}
We show in Fig.\,\ref{fig:pabinit}\,(a)   the electronic pressure of aluminum at standard density and a  temperature of 20\,eV  and 50\,eV (shifted down by 1400\,GPa), computed as a function of the number of orbitals.  We define the convergence by the relative error, in percentage,  of the electronic pressure   $\Delta P= 100 |P-P_0|/P_0$, where $P_0$ is the pressure obtained with the largest number of orbitals. One can see that 128 orbitals per atom are needed at 20\,eV, and 384 at 50\,eV to reach a convergence better than1\% (see numbers in Tables\,\ref{table:occup} and \ref{table:occup2}). Let us recall that at low temperature, for 11 valence electrons, six doubly occupied orbitals \cite{Comment:orbitals}, are sufficient.  The increase from 6 to 128 orbitals per atom represents a computational effort for KSDFT calculations about $20^3$ higher ($35^3$ at 50\,eV).  At high temperature we encounter a paradox situation where a large number of quasi-empty orbitals must be included with a computational cost proportional to the cube of this number (orthogonalization constraint). 
Detailed estimations of the number of orbitals required for a stipulated precision can be found  in recent papers \cite{KARA12b,SJOS14,SJOS15,LUO20}, upon which we will assess our analysis.

The accuracy of the pressure estimation depends on the number of orbitals, because, as soon as the electronic temperature is non-zero, the  distribution of the electronic orbitals is no longer bounded and goes {\it stricto sensu} to infinity. In simulations, involving a finite number of orbitals, the last orbital in energy has a certain level of occupancy, that we will call the last level occupancy (\LLO). The first condition for doing sound simulations is to ensure that the LLO\, is low enough to give converged quantities. As shown in a recent study of MgO in the warm dense regime \cite{SOUB19}, a \LLO\, of $10^{-5}$ is necessary to obtain a better than 0.1\% converged pressure above 100\,eV.  The  convergence $\Delta P$ versus LLO is reported  in Fig.\,\ref{fig:pabinit} (b) and in  Tables\,\ref{table:occup} and \ref{table:occup2}  for 20 and 50\,eV. Data with a LLO smaller than $10^{-6}$ are not represented. The low LLO side of Fig.\,\ref{fig:pabinit} (b), beyond 10$^{-2}$  clearly exhibits  a power law. An empirical  relation between the relative error  in the pressure $\Delta P$ and the LLO $\alpha$ is  $\Delta P=a \alpha^b$, with $a=382$ and $b=0.78$ for 20\,eV (Fig.\,\ref{fig:pabinit} (b), solid red line), and $a=769$, $b=0.78$ for 50\,eV (Fig.\,\ref{fig:pabinit} (b), solid green line). These empirical scalings, that  seem weakly dependent on the temperature, give a good order of magnitude of  the last level occupation  $\alpha$  required for a precision $\Delta P$ 
\begin{equation}
\alpha=\left ({\frac {\Delta P}{a}} \right )^{1/b}.
\label{eq:prec}
\end{equation}

In particular, to get a  precision of 1\% (resp. 0.1\%),  a \LLO\,  of  $5.\:10^{-4}$ (resp. $2.3\:10^{-5}$)  is needed  at 20\,eV (red arrows), and  $2\:10^{-4}$  (resp. $1.\: 10^{-5}$) at 50\,eV (green arrows).  An abrupt application of Eq.\,(\ref{eq:prec}), with parameters for aluminum at 50\,eV  ($a=769$, $b=0.78$, $\Delta P=0.1\%$), predicts a LLO $\simeq \,10^{-5}$ close to the one mentioned in the MgO paper \cite{SOUB19} at 100\,eV. We believe that these scalings are very general and can found for any material and thermodynamic quantities.

The precision of a finite temperature calculation is thus tightly bound to the occupancy level of the highest band in energy involved in the calculation. An estimation of this number is  desirable to calibrate the parameters of a simulation (number of bands, number of atoms) for given computer resources.
\section{Homogeneous electron gas  model}

The electronic state occupancy being given by the Fermi-Dirac statistics, the fundamental parameter  for high temperature KSDFT simulations is not the temperature itself, but rather the Fermi degeneracy defined by $\theta=k_BT/\epsilon_F$. 

For a HEG, the Fermi energy $\epsilon_F$,  expressed in atomic units $e=m=\hbar=1$ reads
\begin{equation}
\epsilon_F=k_BT_F={\left ({\frac{2}{\G}} \right )^{2/3} }    {\frac{(3\pi^2)^{2/3}}{2}} n_V^{2/3},
\label{eq:fermi}
\end{equation}
where $n_V=N_V/V_{at}$ is the electronic density. $V_{at}$ is the atomic volume,  $N_V$ is the number of valence electrons, and \G\, the degeneracy of the electronic state.

 In a KSDFT approach the valence electrons are defined as electrons not belonging to the frozen core of the pseudo-potential. These electrons participate to the global electronic density and can be bound or free. Usually, for aluminum, the outermost three electrons ($3s^2\:3p^1$)  are counted in the valence states, which is enough up to a few eV. But if we want to go at much higher temperatures, 11 electrons ($2s^2\:2p^6\,3s^2\:3p^1$) must be considered, leaving the very deep $1s^2$ states in the core.  To reach extreme conditions, beyond 1\,000~eV, an all-electron description is necessary.

The number of orbitals fulfilling a stipulated  \LLO\, $\alpha$ can be estimated from the HEG properties. In the following, we use the  Fermi-Dirac distribution with a \G\, degeneracy as in the  \Abinit\,   software package
\begin{equation}
\label{eq:FD}
f(\epsilon)={\frac{\G}{ e^{\beta(\epsilon-\mu)}+1  }},
\end{equation}
where $\beta=1/k_\text{B}T$, $\epsilon$ the energy and $\mu$ the chemical potential.
 
  The energy for which the Fermi-Dirac distribution is equal to $\alpha$ is 
\begin{equation}
E_{\alpha}^*=\theta \ln \left[ {\frac{\G}{\alpha}}-1 \right ]+\mu^*,
\label{eq:emax}
\end{equation}
where we have introduced the dimensionless quantities  $\mu^*=\mu/\epsilon_F$ and $E_{\alpha}^*=E_{\alpha}/\epsilon_F$.
$\alpha=10^{-2}, 10^{-3}, ..., 10^{-6}$ is the requested last orbital occupation . 
The chemical potential of the HEG, obtained by comparing  the number of particles at zero temperature with its expression at finite temperature \cite{CHAB98}, reads
\begin{equation}
\mu^*=\theta \:\:I_{1/2}^{-1}\left[ y \right ],
\label{eq:mu}
\end{equation}
where $y={\frac {2}{3}} \theta^{- 3/2}$. 
In this expression, $I_{1/2}^{-1}$ is the inverse of the Fermi integral of order $1/2$.  The chemical potential is equal to the Fermi energy at zero temperature, is zero for $\theta=1$ and becomes negative for large $\theta$. 

The  number of quantum states per atom is 
\begin{eqnarray}
N^*&=&{\frac{\G}{2}} {\frac{2^{3/2}}{3\pi ^2}}V \epsilon^{3/2} \\ \nonumber
	&=&{\frac{\G}{2}} {\frac{2^{3/2}}{3\pi ^2}}V \epsilon_F^{3/2}  (\epsilon/\epsilon_F)^{3/2}  \\ \nonumber
	&=&	N_V {\epsilon^*}^{3/2}. \nonumber
\end{eqnarray}
The number of orbitals is thus
\begin{equation}
N_o=N_V/\G*{\epsilon^*}^{3/2}.
\end{equation}

\begin{figure}[!t]
\begin{center}
\includegraphics[width=8cm]{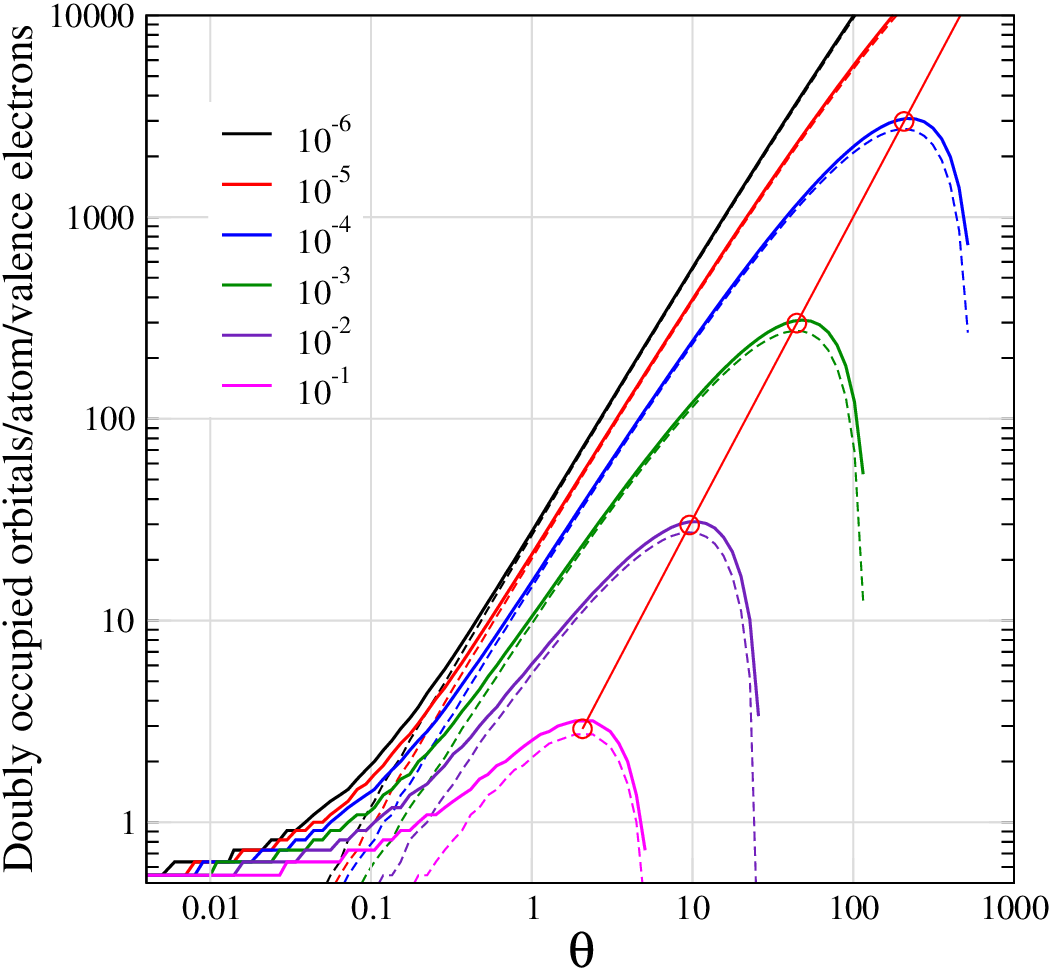}
\caption{(Color online)  Universal curve giving the number of doubly occupied orbitals per atom and per valence electrons in the HEG model  versus degeneracy $\theta$. Colors represent the various stipulated lowest occupations \LLO\, from 10$^{-6}$ (top black) to $10^{-1}$ (bottom magenta). Solid lines: exact calculation (\ref{eq:noexact}) and  dashed lines: high temperature approximation (\ref{eq:noht}). The solid red line with red circles locates the corresponding maxima.}
\label{fig:occup}
\end{center}
\end{figure}

Introducing Eq.\,(\ref{eq:emax}), we obtain the number of doubly occupied states (\G=2) per atom and per valence electrons 
\begin{equation}
\label{eq:noexact}
N_o=\theta^{3/2}  \left ( \ln \left [{\frac{2}{\alpha}} -1\right ]+I_{1/2}^{-1}\left[ {\frac{2}{3}}\theta^{-3/2} \right ] \right )^{3/2}/2 . 
\end{equation}
The total number of doubly occupied orbitals needed for a simulation of $N_\text{at}$ atoms  with Nv valence electrons is  thus $N_\text{tot}=N_o N_\text{at}N_v$.

Using the high temperature  approximation $\beta \mu \approx \ln[y]$ and dropping the 1 in the first logarithm of Eq.\,(\ref{eq:noexact}) yields the simplified expression   
\begin{equation}
\label{eq:noht}
N_o^\text{HT}\approx      \theta^{3/2}   \left ( \ln  \left [ {\frac {4}{3\alpha \theta^{3/2}}} \right ]            \right ) ^{3/2} /2.
\end{equation}
The latter expression predicts about 2-3\% less orbitals at high temperature ($T\gtrsim T_F$) than the exact one but does not require to compute a Fermi integral.  This formulation must not be used below $0.1T_F$. 

The universal curve giving the number of doubly occupied orbitals per atom and per valence electrons  versus reduced temperature $\theta$ is  drawn in Fig.\,\ref{fig:occup}, for increasing \LLO s from $10^{-6}$ (top black) to $10^{-1}$ (bottom magenta).  The constant \LLO\, curves are non-monotonic, but exhibit a maximum and then drop to zero. For each \LLO, there is a maximum temperature $\theta_\text{max}$  beyond which  a bijection between the number of orbitals and temperature cannot  be established. The temperature corresponding to this maximum is well approximated, by taking the temperature derivative of Eq.\,(\ref{eq:noht})
\begin{equation}
\theta_\text{max} =  {\frac {1}{\e}}\left ({\frac {2\G}{3\alpha}} \right )^{2/3},
\label{eq:maxt}
\end{equation}
where $\e=2.71838$ is the usual Neper or Euler number. The maxima correspond  to the maximum doubly occupied orbitals per atom 
\begin{equation}
N_{\alpha}=    {\frac {1}{ \e^{3/2}}}   \left ( {\frac {3}{2}} \right )^{3/2} {\frac {2\G}{3\alpha}} ,
\label{eq:maxN}
\end{equation}
that are shown by open circles in Fig.\,\ref{fig:occup}. 
 These maxima are connected by the relation $N_{\alpha}=(3/2)^{3/2} \theta^{3/2}$, solid red line  in Fig.\,\ref{fig:occup}. 

\begin{figure}[!t]
\begin{center}
\includegraphics[width=7.8cm]{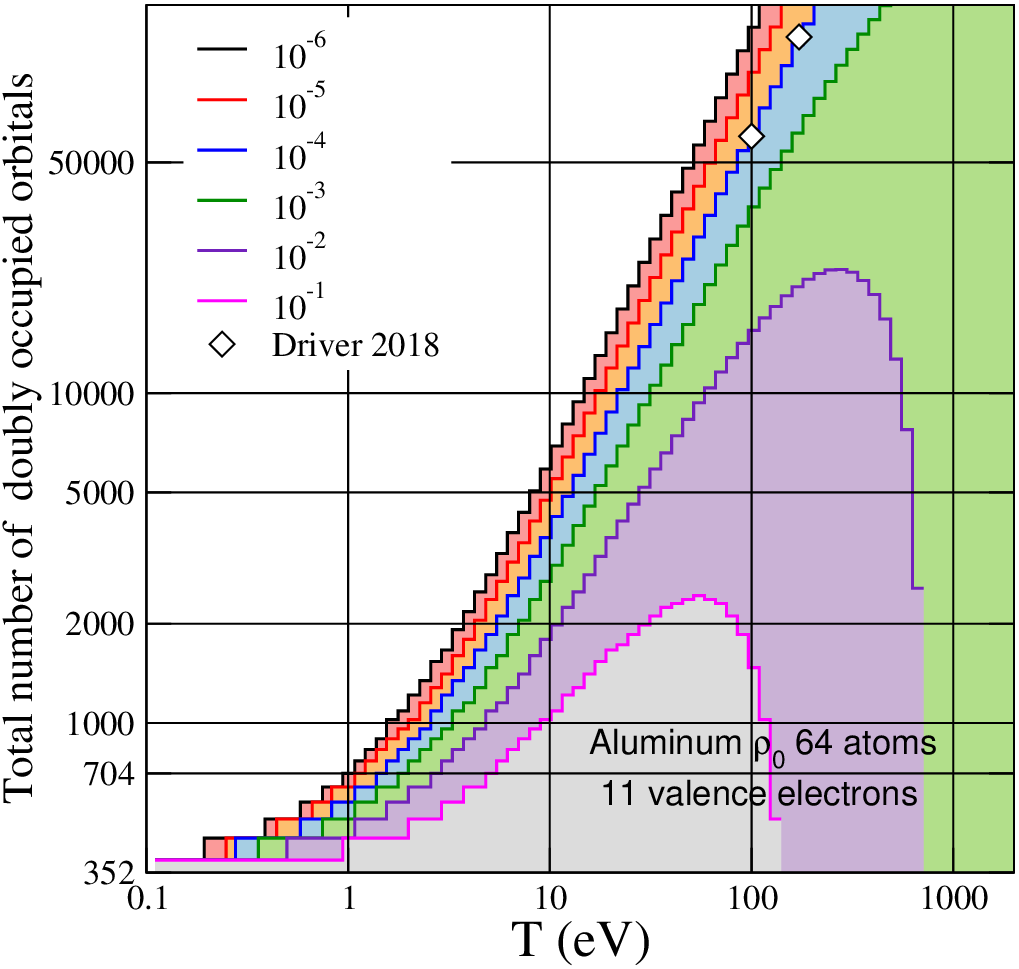}
\caption{(Color online) Same as Fig.~\ref{fig:occup} for a system of 64 aluminum  atoms at standard density  versus temperature ($N_{V}=11$ valence electrons). White diamonds represent the total number of orbitals used by 
Driver \cite{DRIV18} for 8 atoms simulations at 100 and 200~eV, rescaled to 64 atoms.} 
\label{fig:occup2}
\end{center}
\end{figure}

In Fig.\,\ref{fig:occup2} we apply Eq.\,(\ref{eq:noexact}) to a system of 64 aluminum atoms with 11 valence electrons at standard density. At low temperature ($0.1$~eV), we can see that ($64\times 11/2=352$) doubly occupied orbitals are enough to satisfy any level of accuracy. At $ 10$~eV this number grows to $1\,000$ orbitals for a \LLO\, of $10^{-1}$ and  3\,700 orbitals for a  \LLO\, of $10^{-4}$. These numbers are rapidly growing with temperature (as $T^{3/2}$), reaching values of about $50\,000$ orbitals at 100~eV. Lets recall that for aluminum at standard density the Fermi energy is $26.65$~eV for a 11 electrons pseudo-potential.
We have reported in  Fig.\,\ref{fig:occup2} the number of orbitals used

 As a test, we have added in Fig.\,\ref{fig:occup2} two points taken from KSDFT  simulations done by Driver \cite{DRIV18}, on aluminum at 100 and 200~eV and at standard density. For a LLO of $10^{-4}$, up to 76\,000 bands are needed for  a 64 atoms simulation at 100\,eV, in agreement with the predictions of our model  \cite{comment_Soubiran}. To reduce the computational load, 16 and  8 atoms were used at 100\,eV in the above mentioned simulations.

\section{Validation}
 
\subsection{Comparison  with occupations given by \Abinit, }
\begin{figure}[!t]
\begin{center}
\includegraphics[width=8cm]{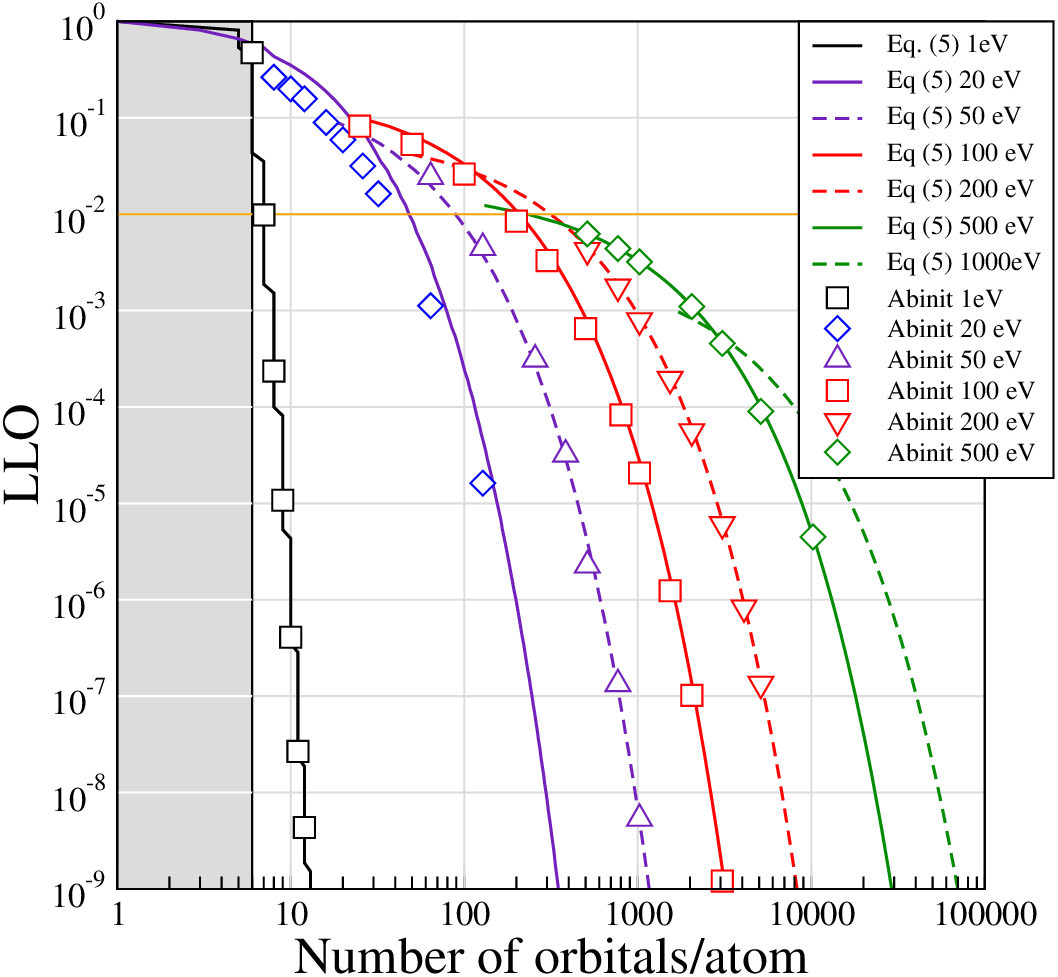}
\caption{(Color online) Last level occupation from \Abinit\,minimizations (symbols) on aluminum compared with  Eq.\,(\ref{eq:noexact}) (solid and dashed lines)  for different temperatures from 0.001 to 1\,000\,eV, at standard density.}
\label{fig:Abinit}
\end{center}
\end{figure}

Fig.\,\ref{fig:Abinit} compares the LLO predicted by the HEG model (Eq.\,(\ref{eq:noexact}), solid and dashed lines)  with the same quantity obtained from an   \Abinit\, electronic minimization (symbols), with an increasing number of orbitals at a given temperature for a four  aluminum atoms system at standard density.  For each  minimization, the occupations are averaged over the $8^3$ k-points. The agreement is good on a wide range of high temperatures. The only difference is observed at low temperature (20\,eV) where the HEG formula clearly overestimates the LLO obtained with \Abinit\,(blue diamonds).

\subsection{Comparison with previous high temperature KSDFT simulations}
\begin{figure}[!t]
\begin{center}
\includegraphics[width=8cm]{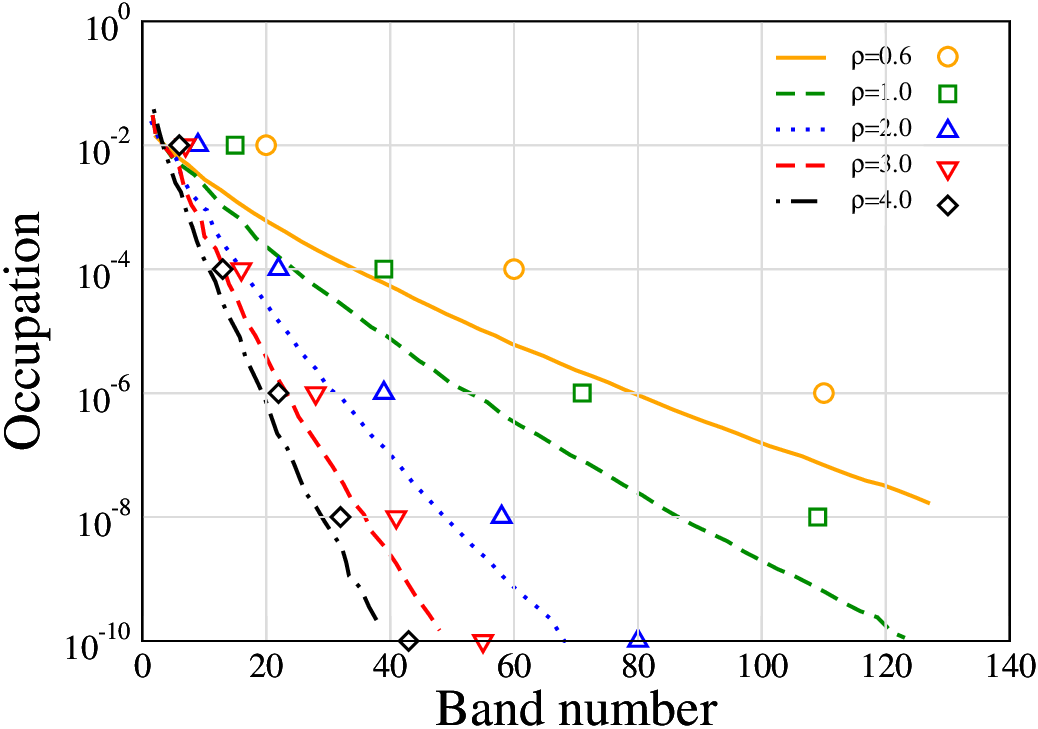}
\caption{(Color online) Average occupations versus band number for lithium at $8.6$~eV and for various densities listed (in \udens). Lines: data from Karasiev \cite{KARA12b}; symbols: our model.  }
\label{fig:Karasiev}
\end{center}
\end{figure}

The first paper to present a careful estimation of population levels, is a study of warm dense lithium\cite{KARA12b}  at various densities and at temperatures up to 100~kK (8.6~eV). We  reported the corresponding data\cite{Comment_Kara}, in Fig.\,\ref{fig:Karasiev} for the various lithium densities (lines). Our prediction is shown by symbols of corresponding colors. We note that at standard density (orange line and circles) the band number is overestimated by about 35\% (e.g. 110 bands instead of 80 for a stipulated occupation of $10^{-6}$), but reduces to 10\% at high density (4\udens, 22 predicted bands for 20 measured in the simulation for  for a stipulated occupation of $10^{-6}$). This is consistent with the previous \Abinit\, minimizations shown in Fig.\,\ref{fig:Abinit}.
We interpret the depopulation of high energy states with the density, as the consequence of the modification of the chemical potential.

We have also reported in Fig.~\ref{fig:Luo}  the number of bands used by Luo {et al.} \cite{LUO20} (Table I of supplemental) to comply a  with a stipulated occupation of $10^{-6}$ for a simulation of  4 aluminium atoms at 2.3\udens (blue circles in Fig.~\ref{fig:Luo} (a)) and a eight silicon atoms at 2\udens\, (green circles in Fig.~\ref{fig:Luo} (b)).  We observe in both cases a very good agreement with Eq.\,(\ref{eq:noexact}) (blue line).

\begin{figure}[!t]
\begin{center}
\includegraphics[width=8cm]{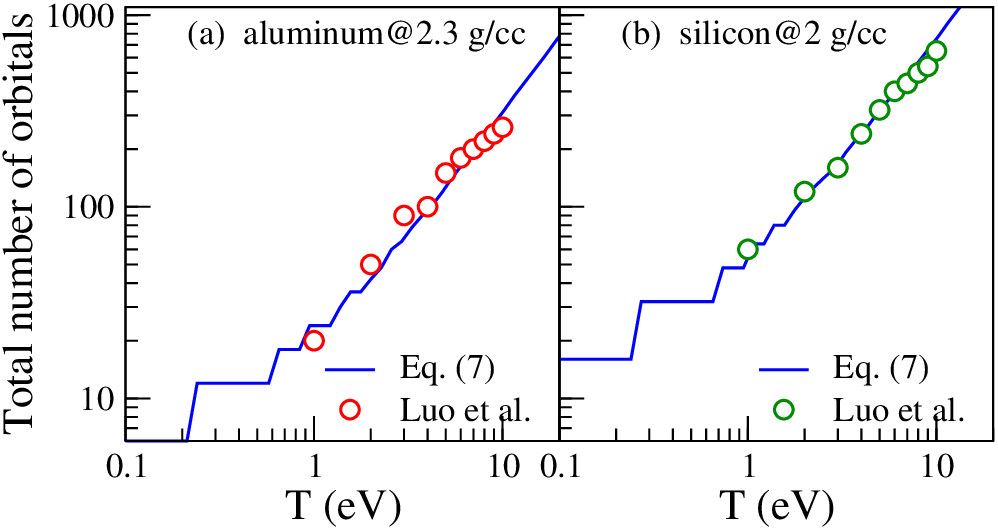}
\caption{(Color online) Comparison of the number of orbitals versus temperature predicted by of Eq.\,(\ref{eq:noexact}) (blue curve)  with Luo \etal\, \cite{LUO20} (Table I of supplemental) for: 
(a)  4 aluminum atoms at 2.3\udens\,(blue circles), 
and (b) eight silicon atoms at 2\udens\, (green circles).  The LLO is $10^{-6}$.}
\label{fig:Luo}
\end{center}
\end{figure} 

Finally, we checked that the predictions of our model (Eq.\,\ref{eq:noexact}) are consistent with the number of orbitals mentioned in various publications \cite{RECO09, LAMB11,SJOS13,SJOS14,SJOS15,DING17} with a slight  trend to overestimate the number of bands at low temperature. For instance, we obtain 192 bands fore a 64 aluminum system  with a LLO=$10^{-3}$ at standard density and at 1\,eV instead of 180, as published by Sjostrom  \cite{SJOS15}, and to be compared to a minimum of 96 bands at zero temperature.

In conclusion of this section, the simple HEG model predicts a number of bands satisfying a stipulated precision in agreement with previous high temperature simulations. At low density and low temperature our model overestimates this number but provides a safe estimate, which can be very large, making KSDFT simulations  very time consuming in some situations.
This calls for models tailored to reduce this number at high temperature as we are going to show with the \ext.

\begin{figure}[!t]
\begin{center}
\includegraphics[width=8cm]{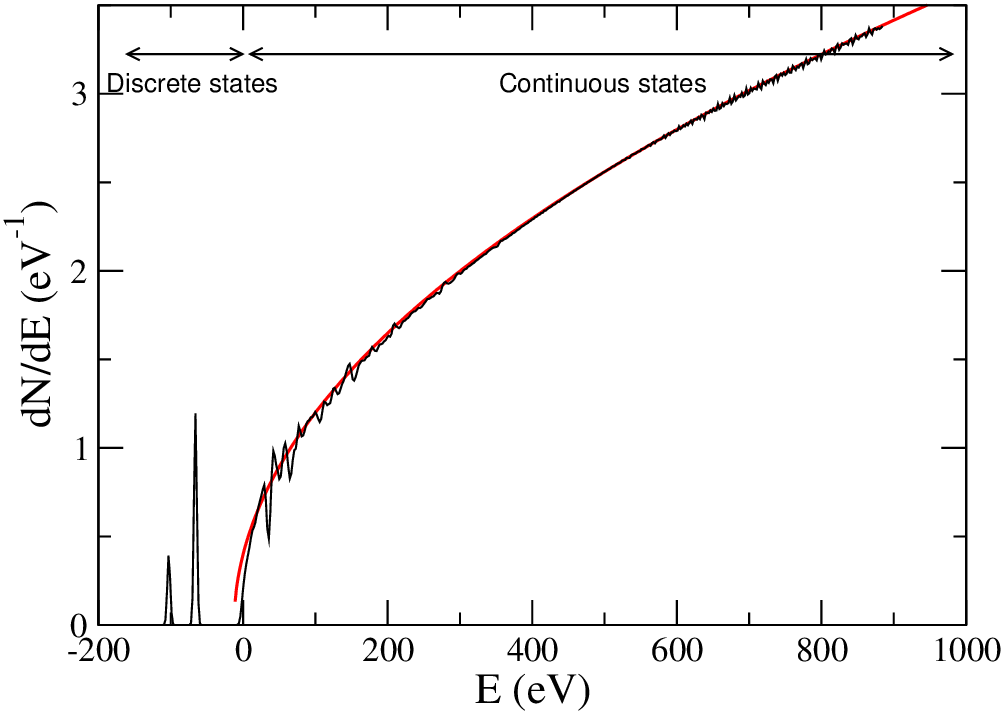}
\caption{(Color online) FCC Al DOS at 20\,eV and standard density computed with 770 $k$-points. In red, the HEG density of states.}
\label{fig:truedos}
\end{center}
\end{figure}

\section{Bypassing the number of orbitals constrain}
\subsection{The extended FPMD method}

In their paper Zhang \etal\, \cite{ZHAN16} suggested to use  the HEG properties to simplify the description of hot dense matter. This connection, between high energy orbitals and HEG  is particularly clear when we consider the density of states (DOS) of a hot system shown in Fig.\,\ref{fig:truedos}.  The DOS of   aluminum at standard density and at 20\,eV, computed  with \Abinit\, and averaged over 770 $k$-points  reveals bounded levels, at negative energies,  merging with a continuum at high energy which closely follows the HEG result 
\begin{equation}
D(\epsilon)={\sqrt{2}\Omega \over \pi^2} \sqrt{\epsilon-U_0},
\label{eq:dos}
\end{equation}
where $\Omega$ is the volume and   $U_0$ a shift in the energy, to be determined.

At finite temperature,  the evaluation of thermodynamic quantities (density, energy, entropy) can be broken into two parts: a discrete part accounting for $N_c$ discrete levels  and a quasi-continuous part, corresponding to  densely distributed atomic states,  through the DOS  $g(\epsilon)$. The energy, for example, can be written as
\begin{equation}
E=-\sum_{i=1}^{N_c} f(\epsilon_i) \left < \psi_i \left | \nabla^2 \right | \psi_i\right > + \int_{E_C}^{\infty} f(\epsilon)g(\epsilon)\epsilon d\epsilon,
\end{equation}
where $N_c$ is the number of considered eigenstates states $\psi_i$ with occupation $f$. The occupation $f(\epsilon_c)$  is nothing else than our previously introduced \LLO. Fig.\,\ref{fig:truedos} suggests, to use the HEG density of states, beyond asome cutoff energy. For the electronic density, by example, we end up with the following formulation
\begin{eqnarray}
n({\bf r})&=&2\sum_{i=1}^{N_c} f(\epsilon_i) \left | \psi_i({\bf r})  \right |^2 -{1 \over \Omega} \int_{E_C}^{\infty} f(\epsilon) D(\epsilon)d\epsilon\\
	&=& n({\bf r})_{KS} +n_0,
	\label{eq:dens}
\end{eqnarray}
where the constant density $n_0$  is  given by
\begin{equation}
n_0= -{\frac{2 \sqrt{2}} {\pi^2\beta^{3/2}}} I_{1/2}^\text{inc}(\eta, x_c).
\label{eq:dens3}
\end{equation}
$I_{1/2}^\text{c}(\eta,x_c)$ is the incomplete Fermi integral of index $1/2$ of argument $\eta=\beta \mu$ and lower bound $b=\epsilon_c$, whose general definition is 
\begin{equation}
I_{1/2}^{\text{inc}}(\eta,b)=\int_b^{\infty} {\frac {x^{1/2}} {e^{x-\eta}+1}}dx.
\end{equation}
This integral,  that can be very precisely computed, is equivalent to an infinitely small LLO. Zhang {\it el al.} \cite{ZHAN16} have shown that a non negligible part of the precision lies between $10^{-4}$ and $10^{-6}$ LLO.
It must be emphasized that $N_c$ is now fully decoupled from the stipulated precision since the {\it missing} density, pressure or energy can be exactly computed.

\begin{table}[t]
\caption{Comparison with Zhang's paper for hot electrons at 408\,eV in a cold FCC lattice of aluminum at $\rho=2.7\udens$. For each quantity the first line is Zhang's results and the second line, in italic, our result.}
\begin{center}
\begin{tabular}{c|c|c}
\hline \hline
		&Reference	& \extshort				\\

\hline
		&FT-DFT		& Calc. 2		\\
\hline
$\mu$	&-64.419		&-64.512		\\
(Ha)		&{\it -65.410}	&{\it -64.520}	\\
		&			&			\\
E		&-1\,097.33		&-1\,098.41	\\
(Ha)		&{\it -1\,097.09}	&{\it -1\,097.74}	\\
 		&			&			\\
-TS		&-1\,202.777	&-1\,209.604	\\
(Ha)		&{\it -1\,201.236}	&{\it -1\,208.122}	\\
		&			&			\\
P		&4.11	10$^4$	&4.21	10$^4$	\\
(GPa)	&{\it 4.12 10$^4$}	&{\it 4.22 10$^4$}	\\

\hline\hline
\end{tabular}
\end{center}
\label{table:compare}
\end{table}%

\subsection{Validation of implementation}
We implemented the  \ext\, into the \Abinit\,software package, with the projector augmented wave (PAW) method. Depending on the temperature, different atomic datas relatively to the electronic temperature  were considered. From T=$0$\,eV to T=$50$\,eV, we used a PAW-LDA small core (11 valence electrons) pseudo-potential generated by N. A. W. Holzwarth with ATOMPAW software \cite{HOLZ01}. From T=$50$\,eV to T=$500$\,eV, we used a PAW-GGA pseudo-potential with a smaller core, also used by K. P. Driver \cite{DRIV18} on his aluminum KSDFT/PIMC computations. For temperatures higher than T=$500$\,eV,  an ultrasoft all-electrons pseudo-potential generated by V.~Recoules  with ATOMPAW, was necessary. 

For all pseudo-potentials we used a cutoff between 50 and 100\,Ha, after checking, for each temperature, the convergence of the pressure. The core radius was varied from 1.6\,Bohr, at low temperature, to 0.6\,Bohr, 	at high temperature, to prevent any significant PAW spheres  overlapping during molecular dynamics.
For the exchange-correlation functional, we used the local density approximation\cite{CEPE80} up to 200\,eV, and the generalized gradient approximation Perdew-Burke-Ernzerhof (PBE)\cite{PERD96} beyond. All simulations, from 64 to 8 atoms were performed at the $\Gamma$ point.

To check our implementation, we compare in Table \ref{table:compare} the thermodynamic quantities (chemical potential $\mu$, energy $E$, entropy $-TS$ and pressure $P$) for hot electrons at 408\,eV in a cold FCC lattice of aluminum at $\rho=2.7\udens$ given by Zhang \etal\, \cite{Comment_Zhang} with our evaluation. For this particular case, we used the  settings given in Zhang's paper: a cutoff of 250\,Ha, the GGA-PBE exchange correlation, a cutoff radius of 0.6\,Bohr, and an all-electrons PAW pseudo-potential. Our calculation is using the analytical expression (\ref{eq:dens3}) which is equivalent to {\tt calc.\,2} evaluation. We note an excellent agreement with all quantities, better than 0.1\%. The same calculation with 11 electrons  in the pseudo-potential and frozen 1s core electrons would have given 5\% less pressure, signaling  the onset of the ionization of the core electrons in this regime.

To test the efficiency of the method on a wide range of temperatures, we computed the pressure along the aluminum standard density  isochore, from 0.1 to 2\,000\,eV.  Results are gathered in Table\,\ref{table:isochore}, in which  the first column indicates the temperature in eV and the second one, the number of orbitals per atom stipulated by Eq.\,(\ref{eq:noexact}) satisfying  a LLO of 10$^{-4}$. The next column (4) gives the number of orbitals per atom we used in our implementation of the \ext\, which is well below the previous one, and corresponds to a  LLO in the vicinity of 10$^{-2}$. The comparison of our estimation (column 6) with Drivers's results \cite{DRIV18} (column 5) demonstrates an excellent agreement, with an accuracy of about 1\%, for a much lower computational effort, particularly beyond 200\,eV, where classical KSDFT simulations are extremely time-consuming and are replaced by PIMC simulations. The maximum deviation occurs at 1\,eV, reaching 1.5\% and is related to the interpolation of Driver's data, computed at rounded values in K (10\,000, 20\,000, 50\,000 and 1000\,000\,K).

We stress, that even well below 100\,eV, the \ext\,  is roughly one order of magnitude faster than the corresponding KSDFT calculation.

\begin{table}[t]
\caption{Total pressures obtained by the \ext, along the aluminum isochore $\rho_0$. $N_o$ is the prescription given by Eq.\,(\ref{eq:noexact}) for an LLO of 10$^{-4}$ and  $N_o^\text{\extveryshort}$ the number of orbitals used in the calculation. P$_\text{Driver}$ values are interpolation of pressures of  Driver~{\it et al.}  \cite{DRIV18} to the temperatures shown. The last column indicates the precision $100 |P_\text{\extveryshort}- P_\text{Driver}|/P_\text{Driver}$.}
\begin{center}
\begin{tabular}{|c|c|c|c|c|c|c|c|}
\hline 
T	&	$N_o$/at	&  $N_o^\text{Ext}$/at	& 	$N_\text{at}$	&	P$_\text{\extveryshort}$& P$_\text{Driver}$& error		\\
eV	&	Eq.\,(\ref{eq:noexact})	& \extshort	&				&	GPa		&GPa	& \%	\\
\hline
\hline
0.1		&	6		&		6		&		64	&		6.0		&	-		&	-\\
1		&	11		&		11		&		64	&		39.1		&	39.7		&	1.4\\
2		&	11		&		11		&		64	&		70.5		&	71.2		&	1.0\\
10		&	49		&		16		&		64	&		334.7	&	337.2	&	0.7\\
20		&	115		&		32		&		32	&		744.3	&	753.3	&	1.2\\
100		&	814		&		64		&		16	&		7681		&	7693		&	0.2\\
200		&	1\,837	&		128		&		16	&		19\,298	&	19400	&	0.5\\
500		&	5\,126	&		256		&		8	&		59\,907	&	59\,952	&	0.1\\
1\,000	&	10\,428	&		256		&		8	&		129\,170	&	130\,042	&	0.6\\
2\,000	&	19\,332	&		256		&		8	&		264\,018	&	266\,235	&	0.8\\
\hline
\end{tabular}
\end{center}
\label{table:isochore}
\end{table}%

\section{Conclusion}
We have provided a quantitative estimation of the number of orbitals  needed to reach a given level of precision at any temperature for a KSDFT calculation. We have shown that, for a fixed precision, this number increases dramatically with the temperature, making extremely time-consuming KSDFT simulations of matter at standard density and below, beyond a few tens of eVs. We have then implemented the \ext\, in the \Abinit\, software package and shown how the introduction of the homogeneous electron gas density properties allows to correct poorly converged Kohn-Sham calculations with small number of orbitals, allowing to reach keVs temperatures straightforwardly.

\section*{Acknowledgments}
Vanina Recoules, Francois Soubiran and Burkhard Militzer are warmly acknowledged for stimulating discussions and for providing data and pseudo-potentials.

The data on aluminum isochore that support the findings of this study are available in the Supplemental of Driver's paper \cite{DRIV18}, and other data are given in the Tables.


\end{document}